\documentclass[prl,twocolumn,floats,floatfix,showpacs,nofootinbib]{revtex4}
\usepackage{graphicx}
\usepackage{dcolumn}
\usepackage{bm}
\usepackage{graphics}
\usepackage{slashed}
\usepackage{amssymb}
\usepackage{natbib}
\usepackage{amsmath}
\usepackage{url}
\usepackage{color}
\newcommand{\bea}{\begin{eqnarray}}
\newcommand{\ena}{\end{eqnarray}}
\newcommand{\bean}{\begin{eqnarray*}}
\newcommand{\enan}{\end{eqnarray*}}

\begin{document}

\title{A new symmetry of the relativistic wave equation}

\author{F.T. Falciano\footnote{ftovar@cbpf.br}, E. Goulart\footnote{egoulart@cbpf.br}}

\affiliation{Instituto de Cosmologia Relatividade Astrofisica ICRA -
CBPF\\ Rua Dr. Xavier Sigaud, 150, CEP 22290-180, Rio de Janeiro,
Brazil}

\date{\today}

\begin{abstract}
In this paper we show that there exists a new symmetry in the relativistic wave equation for a scalar field in arbitrary dimensions. 
This symmetry is related to redefinitions of the metric tensor which implement a map between non-equivalent manifolds. 
It is possible to interpret these transformations as a generalization of the conformal transformations. 
In addition, one can show that this set of manifolds together with the transformation connecting its metrics forms a group. 
As long as the scalar field dynamics is invariant under these transformations, there immediately appears an ambiguity concerning the definition of the underlying background geometry.
\end{abstract}


\maketitle


In this letter we will show that there exists a novel internal symmetry related to the underlying metrical structure in which a relativistic field propagates. It allows a continuous infinity of active transformations of the metric tensor and, in a sense, is complementary to the usual Lorentz and gauge symmetries. As it is well known, Lorentz symmetry represents translational and rotational invariance in the Minkowski spacetime while gauge symmetries result from the invariance of the equations of motion under certain internal groups of transformations \cite{Wig}-\cite{velt}. Our result reveals a third class of symmetry: invariance under a specific map between non-equivalent manifolds. We will show that the set of allowed transformations constitutes a group and characterizes a class of distinct metric tensors.

Typically, the equation of motion of any field,  in particular of the scalar field $\varphi$, is an intricate amalgam between the underlying metric structure $g_{\mu\nu}$, derivatives of the fields and some field's functional (see, for instance \cite{Leite, wald}). Apart from the context of the Theory of General Relativity, the spacetime structure has to be previously and completely characterized, for instance by giving the metric tensor in a riemannian manifold. This condition is necessary so as to guarantee a well defined dynamics. Within this scenario, the spacetime should be understood as an a priori structure by and over which fields propagate without disturbing its properties. To simplify our reasoning we disregard any gravitational effect and consider the background as given by the Minkowski metric. We shall come back to some considerations related to General Relativity at the end.

In general, each theory presents its external and internal symmetries. Besides being associated with conservation laws and possible internal gauge freedoms, they also play a fundamental role in the very characterization of the elementary particles. The Poincar\' e group distinguishes a privileged class of observers from which all the labels attached to the building blocks of matter are specified (mass, spin, charge, etc) \cite{Wein}. Note, however, that in the framework of conservation laws the metric is taken as an \textit{ad hoc} external object. Even in Noether' s theorem one generally studies the invariance of the action with respect to coordinates and dynamical fields variations, but always keeping the metric fix.

Accordingly, it seems that to a greater or lesser extent all physical results depend on the previously assumed spacetime structure. Nevertheless, we will show that there is a certain degree of arbitrariness in the characterization of the underlying background geometry that was not noted before. Indeed, starting from a very simple scalar field equation in the flat spacetime we demonstrate that its corresponding dynamics is invariant under a specific transformation of the background metric. Hence, the field dynamics does not distinguish which is the spacetime structure within this class of non-equivalent curved metrics. In effect, if the scalar field $\varphi$ is a solution of the Klein-Gordon equation in Minkowski spacetime, then it will also be a solution of the Klein-Gordon equation in each and every curved metric of an infinite family of metrics. The unveiling of this new internal symmetry raises new profound conceptual problems related to quantization and may be viewed also as a mechanism to generate new solutions in curved spacetimes.

Let us start by considering a flat Minkowski spacetime $\cal M$ in $D=1+d$ dimensions with metric tensor $\gamma_{\mu\nu}$ in an arbitrary coordinate system. The simplest possible relativistic wave equation in $\cal M$ is that of a neutral massless spin-zero particle, which is described by the second order, linear, hyperbolic PDE \cite{cou}
\begin{equation}\label{wave} 
 \square \varphi\equiv\frac{1}{\sqrt{-\gamma}}\partial_{\mu}\left(\sqrt{-\gamma}\gamma^{\mu\nu}\partial_{\nu}\varphi\right)=0 \quad ,
\end{equation}
with $\varphi$ real and $\gamma\equiv det(\gamma_{\mu\nu})$. Let us consider a symmetric tensor field $q^{\mu\nu}$ in $\cal M$ defined as
\begin{equation}\label{q}
q^{\mu\nu}\equiv A\ \gamma^{\mu\nu}+B\ \gamma^{\mu\alpha}\gamma^{\nu\beta}\partial_{\alpha}\varphi\ \partial_{\beta}\varphi \quad,
\end{equation}
where $A$ and $B$ are two arbitrary continuous and differentiable real functions. Denoting the canonical kinetic term as $w\equiv\gamma^{\mu\nu}\partial_{\mu}\varphi\ \partial_{\nu}\varphi$, the following identity holds
\[
q^{\mu\nu}\partial_{\nu}\varphi=(A+Bw)\gamma^{\mu\nu}\partial_{\nu}\varphi \; .
\]
If $q^{\mu\nu}$ is non-degenerate there exists a new tensor\footnote{Note that $q^{-1}_{\mu\nu}\neq q^{\alpha\beta}\gamma_{\alpha \mu}\gamma_{\beta \nu}$.} $q^{-1}_{\mu\nu}$ such that $q^{\mu\alpha}q^{-1}_{\alpha\nu}=\delta^{\mu}_{\phantom a\nu}$. In general, the inverse of an object $g^{\mu \nu}=\eta^{\mu \nu}+h^{\mu \nu}$, with arbitrary $h^{\mu\nu}$, is given as an infinite series\footnote{The inverse of $g^{\mu \nu}=\eta^{\mu \nu}+h^{\mu \nu}$, with arbitrary $h^{\mu\nu}$, is given as $g_{\mu \nu}=\eta_{\mu \nu}-h_{\mu \nu}+h_{\mu}{}^\alpha h_{\alpha \nu}+\ldots$}. Notwithstanding, due to algebraic properties encoded in $q^{\mu\nu}$, its inverse is simply
\begin{equation}
q^{-1}_{\mu\nu}=\frac{1}{A}\gamma_{\mu\nu}-\frac{B}{A(A+Bw)}\partial_{\mu}\varphi\partial_{\nu}\varphi \quad .
\end{equation}

The determinant of $q^{-1}_{\mu\nu}$ in $D=1+d$ dimensions may be easily calculated using Sylvester's determinant theorem which states that if $\bold M$ and $\bold N$ are respectively $p\times q$ and $q \times p$ matrices, then
\begin{equation*}
det\left( \bold{1}_{p}+\bold{M}\bold{N}\right)=det\left(\bold{1}_{q}+\bold{N}\bold{M}\right) \quad,
\end{equation*}
where $\bold{1}_{r}$ is the identity matrix of order $r$. A direct calculation yields the expression
\begin{equation*}
\sqrt{-det\left(q^{-1}_{\mu\nu}\right)} =\sqrt{-\gamma}\left[A^{-d/2}(A+Bw)^{-1/2}\right]\quad .
\end{equation*}

Hereafter, we shall envisage the tensor field $q^{\mu\nu}$ as defining a metric on a new riemannian manifold $\widehat{\cal{M}}$. In addition, we shall consider if it is possible to choose $A$ and $B$ such that any given solution of the massless Klein-Gordon equation (\ref{wave}) with metric $\gamma^{\mu\nu}$ is also a solution of the massless Klein-Gordon equation in the $q^{\mu\nu}$ metric spacetime.

Curiously enough, by choosing $B=(A^{d}-A)/w$ with $A(x)$ still completely arbitrary, there is such a degeneracy in the description of the metrical structure. Consequently, the equation of motion that describes a real massless scalar field in a D-dimensional Minkowski spacetime is invariant under all possible transformations of the metric $\gamma^{\mu\nu}\rightarrow\  q^{\mu\nu} $ with
\begin{equation}\label{qA}
q^{\mu\nu}=A\left\{\gamma^{\mu\nu}+\frac{A^{d-1}-1}{w} \gamma^{\mu\alpha}\gamma^{\nu\beta}\partial_{\alpha}{\varphi} \partial_{\beta}{\varphi}\right\}.
\end{equation}

Thus, if $\varphi$ is a solution of equation (\ref{wave}), then it also satisfies
\[
\frac{1}{\sqrt{-det\left(q^{-1}_{\mu\nu}\right)}}\ \partial_{\mu}\left(\sqrt{-det\left(q^{-1}_{\mu\nu}\right)}\ q^{\mu\nu}\partial_{\nu}\varphi\right)=0 \quad .
\]

For a fixed solution $\varphi(x)$, transformation (\ref{qA}) is completely characterized by the function $A(x)$. Therefore, we shall label each metric with a subscript $q^{\mu\nu}_{(A)}$. Furthermore, one should view $q^{\mu\nu}_{(A)}$ as a metric tensor that defines a new spacetime. In this new manifold, every tensor should have its indices raised and lowered by the metric $q^{\mu\nu}_{(A)}$. In particular, the inverse of the metric which is given by the relation $q_{(A)}^{\mu\alpha} q_{(A)\alpha\nu}=\delta^\mu{}_\nu$ is simply $q_{\mu\nu}^{-1}$, i.e. $q_{(A)\mu\nu}=q_{\mu\nu}^{-1}$. 

This nontrivial result brings into light an interesting ambiguity concerning the very definition of the background metric. In fact, due to the described symmetry, it seems that it is impossible to distinguish between these metrics by studying the dynamics of the $\varphi$ field. The scalar field has the same evolution in all these different spacetimes, i.e given an arbitrary solution $\varphi$ of the wave equation (\ref{wave}), there exists an infinite class of curved spacetimes in which $\varphi$ is also a solution. 

At this point, we should mention that the above internal symmetry in the dynamical equation should not be mistaken by general covariance. Indeed, one can easily show that the metric $\gamma^{\mu\nu}$ cannot be mapped into $q^{\mu\nu}_{(A)}$ by a simple coordinate transformation. First note that in its definition, equation (\ref{qA}), there is a global factor $A(x)$ which plays a role similar to a conformal factor. Thus, it cannot be a coordinate transformation if $A(x)$ is a non-constant function. Notwithstanding, even when $A$ is constant, it is still not a coordinate transformation. We can convince ourselves by noting that in this case the metric assumes the form of a Gordon-like metric \cite{gordon, nov1}. In analogue models of gravitation \cite{matt, nov2}, a Gordon metric is of the form
\[
g^{\mu \nu}=a_{0}\eta^{\mu \nu}+b_{0} v^{\mu}v^{\nu}\quad ,
\]
with $a_{0}=const$, $b_{0}$ defined in terms of the electromagnetic properties of the medium and $v^{\mu}$ a normalized vector in the Minkowskian inner product. This is precisely the same form of $q^{\mu\nu}_{(A)}$ when $A$ is constant. Conclusively, we can also show that $q^{\mu\nu}_{(A)}$ is not Minkowski spacetime written in a different coordinate system by displaying a specific example. Consider a static spherical symmetric solution of the Klein-Gordon equation. This solution is of the form $\varphi=\lambda r^{-1}$ with $\lambda$ a real constant. Pluging this solution in equation (\ref{qA}) one can show that the curvature tensor associated with $q^{\mu\nu}_{(A)}$ is non-zero which is a result that cannot be accomplished by a coordinate transformation.

While covariance is connected with different coordinate coverings in the same manifold $\cal{M}$, the present symmetry defines a set $\left\{\widehat{\cal{M}}\right\}$ of different manifolds. Each one of them is parametrized by a realization of the function $A(x)$. Thus, if we are constrained to make experiments only with the massless scalar field $\varphi$, we are not able to distinguish in which particular realization the field propagates, at least at the classical level. 

Note that if $d=1$ the transformation (\ref{qA}) reduces to a conformal transformation (see \cite{conf, Kast} for a detailed discussion). This is the unique space dimension where the transformation does not depend on the particular solution $\varphi$. Hence, the well known conformal invariance of the wave equation in (1+1) dimensions can be viewed as a particular case of transformation (\ref{qA}). In this sense, this new symmetry transformation is a generalization of the conformal transformation.

Our analysis has focused in the invariance of the equation of motion (\ref{wave}) under metric transformations and, as it is well known, symmetries of the equation of motion do not imply symmetries in the action. However, it is straightforward to show that transformation (\ref{qA}) is also a symmetry of the action. Indeed, the action integral
\begin{equation}
S=\int\gamma^{\mu\nu}\partial_{\mu}\varphi\ \partial_{\nu}\varphi\sqrt{-\gamma}d^{4}x
\end{equation}
is invariant under the map
\begin{equation}
 \gamma^{\mu\nu}\ \rightarrow \ q_{(A)}^{\mu\nu}\quad\quad \sqrt{-\gamma}\ \rightarrow \ \sqrt{-q_{(A)}}
\end{equation}
with $q_{(A)}\equiv det\, q_{(A)\,\mu\nu}$. 

In a Taylor expansion around a fiducial point $x_{0}$, the function $A(x)$ is characterized by an infinite number of parameters. Consequently, the continuous symmetry of the action  is also characterized by an infinite number of parameters \cite{Had}. 

In each curved manifold belonging to $\widehat{\cal M}$, there is a conservation law that is associated with the propagation of the scalar field in the $q_{(A)}^{\mu\nu}$ metric. In each riemannian manifold defined by $q_{(A)}^{\mu\nu}$, one has the metricity condition
\begin{equation}
\nabla_{(A)}^{\, \mu}q_{(A)}^{\alpha\beta}=0 \quad,
\end{equation}
where the subscript $_{(A)}$ in the covariant derivative operator indicates that it is constructed with the metric $q_{(A)}^{\alpha\beta}$. Defining the tensor
\begin{equation}
T_{(A)}^{\mu\nu}\equiv  q_{(A)}^{\mu\alpha}q_{(A)}^{\nu\beta}\partial_{\alpha}\varphi\ \partial_{\beta}\varphi-\frac12 q_{(A)}^{\mu\nu} \, q_{(A)}^{\alpha\beta}\partial_{\alpha}\varphi\ \partial_{\beta}\varphi
\end{equation}
it is immediate to check that it is divergenceless with respect to $q_{(A)}^{\mu\nu}$, i.e.
\begin{equation}
\nabla_{(A)\,\mu}\, T_{(A)}^{\mu\nu}=0\quad .
\end{equation}

Another interesting property of the symmetry transformation (\ref{qA}) is that together with the set of differential manifolds $\left\{\widehat{\cal{M}}\right\}$ they form a group for each and every solution $\varphi$. To simplify notation, from now on we define the transformation symbol ${\cal T}_a$ associated with the function $a(x)$. An application of ${\cal T}_a$ in the $\gamma^{\mu\nu}$ metric is such that 
\begin{equation}
{\cal T}_a\ [\gamma^{\mu\nu}]\equiv q^{\mu\nu}_{(a)}
\end{equation}
with $q^{\mu\nu}_{(a)}$ defined by the rule (\ref{qA}). According to our previous discussion the transformation symbol relates two non-equivalent manifolds. Let us show that a successive transformation ${\cal T}_b$ associated with the function $b(x)$ yields again a tensor of the form $q^{\mu\nu}_{(c)}$. We have
\begin{equation}
{\cal T}_b\ [{\cal T}_a\ [\gamma^{\mu\nu}]]={\cal T}_b\ [q^{\mu\nu}_{(a)}]\quad.
\end{equation}
Replacing all $\gamma^{\mu\nu}$ by $q^{\mu\nu}_{(a)}$ into (\ref{qA}) one immediately obtains
\[
{\cal T}_b\ [q^{\mu\nu}_{(a)}]=b\left\{q^{\mu\nu}_{(a)}+\frac{b^{d-1}-1}{w_{(a)}} q^{\mu\alpha}_{(a)}q^{\nu\beta}_{(a)}\partial_{\alpha}{\varphi} \partial_{\beta}{\varphi}\right\}\quad,
\]
where $w_{(a)}\equiv q^{\mu\nu}_{(a)}\partial_{\mu}\varphi\partial_{\nu}\varphi$. A direct calculation using explicitly $q^{\mu\nu}_{(a)}$ in terms of $\gamma_{\mu\nu}$ gives us
\[
{\cal T}_b\ [q^{\mu\nu}_{(a)}]=c\left\{\gamma^{\mu\nu}+\frac{c^{d-1}-1}{w}\gamma^{\mu\alpha}\gamma^{\nu\beta} \partial_{\alpha}{\varphi} \partial_{\beta}{\varphi}\right\}\quad,
\]
with $c=b.a$. Therefore, the composition of two successive transformation, $\gamma \xrightarrow{{\cal_T}_a} q_{(a)} \xrightarrow{{\cal_T}_b} q_{(c)}$ equals a single transformation $\gamma \xrightarrow{{\cal_T}_{ba}} q_{(ba)}$, i.e. $q_{(c)}=q_{(ba)}$. Thus, the contraction of any two objects of the form  (\ref{qA}) with $\gamma_{\mu\nu}$ is again an object of the same type. A carefull inspection of the transformations ${\cal T}_a$ reveals that all the usual group properties are verified, i.e.
\begin{itemize}
\item[i)]Identity
\[
{\cal T}_1\circ {\cal T}_a={\cal T}_a\circ {\cal T}_1={\cal T}_a
\]
\item[ii)] Inverse 
\[
{\cal T}^{-1}_a={\cal T}_{(a^{-1})}
\]
\item[iii)] Closure
\[
{\cal T}_b\circ {\cal T}_a= {\cal T}_{(b\, a)}
\]
\item[iv)] Associativity
\begin{eqnarray*}
&&{\cal T}_c\circ \Big({\cal T}_b\circ {\cal T}_a\Big)={\cal T}_c\circ {\cal T}_{(ba)}={\cal T}_{(cba)}\quad\\
&&\qquad ={\cal T}_{(cb)}\circ {\cal T}_a=\Big({\cal T}_c\circ {\cal T}_b\Big)\circ {\cal T}_a
\end{eqnarray*}
\end{itemize}
It is worth noting that, since each transformation is characterized by a real function, we are dealing with an infinite parameter abelian Lie group. 

Actually, there is two possible ways to interpret transformation (\ref{qA}). First, one can view equation (\ref{qA}) as a rule to define a new metric of a different manifold. Hence, it is a prescription to map different riemannian manifolds as we have been considering so far. However, there is another possibility that is to consider equation (\ref{qA}) as a definition of a family of tensors in the same spacetime, i.e. the only metric is the Minkowski metric. Thus, all tensors should be raised and lowered with $\gamma^{\mu \nu}$. In this second approach, the family of tensors defined by (\ref{qA}) is a realization of the group described above and the composition between two of its elements is made through the metric $\gamma^{\mu \nu}$. Indeed, the product of two elements gives
\begin{equation}
q^{\mu\alpha}_{(a)}\, \gamma_{\alpha\beta}\, q^{\nu\beta}_{(b)}=q^{\mu\nu}_{(ab)}\quad.
\end{equation}

Therefore, in this realization the identity element is the $\gamma^{\mu\nu}$ metric and the inverse of $q^{\mu\nu}_{(a)}$ is given by $q^{\alpha \beta}_{(a^{-1})}\, \gamma_{\mu \alpha}\, \gamma_{\nu \beta}$. In order to distinguish between these two approaches we may refer to them as an active or passive transformation. The active transformation defines different metrics and maps different riemannian manifolds while the passive transformation defines a family of tensors in the same manifold.

Finally, we would like to comment on how it is possible to generalize our results. First, instead of considering a massless free scalar field, one could consider a mass term $m^{2}\varphi^{2}$ or include self-interaction through a potential $V(\varphi)$ in the lagrangian. In both cases, all the above results are preserved if in addition to the transformation of the metric one also transforms the mass and the potential of the scalar field. Thus, if the field $\varphi$ is a solution of the Klein-Gordon equation with mass $m$ and potential $V(\varphi)$ in the $\gamma^{\mu\nu}$ spacetime, then it is also a solution of the Klein-Gordon equation with
\begin{equation}
m^{2}\rightarrow A^d m^2\quad\quad V_{A}(\varphi)\rightarrow A^dV(\varphi)
\end{equation}
 in the $q_{(A)}^{\mu\nu}$ spacetime. Thus, the map endows a position dependent mass term or in the case of a constant function $A$ it can simply renormalize the mass of the scalar field.

Concerning General Relativity, one could examine how this ambiguity in the definition of the metrical structure can be sustained in a gravitational scenario. Suppose there is a single scalar field with energy density strong enough so that it is imperative to consider how it deforms the spacetime structure. The solution that satisfies simultaneously Einstein's equations and the Klein-Gordon equation is a pair of fields $(g_{\mu\nu}\, ,\varphi)$. Thus, it seems that GR would imply a unique characterization of the metric as far as the propagation of $\varphi$ is concerned. 

However, even in this case, there is still a degeneracy in the metrical structure of the spacetime inasmuch we don't have a direct access to the metric $g_{\mu\nu}$. Supposing that we can only survey the spacetime by studying the behavior of its matter content, i.e. by analyzing the evolution of the scalar field $\varphi$, then we will still have an ambiguity in the definition of its metrical structure. 

In fact, as long as the Klein-Gordon equation still maintains its internal symmetry, one can define an infinite family of spacetimes by replacing the Minkowski metric in equation (\ref{qA}) by the $g_{\mu\nu}$ metric coming from Einstein's equations. Thus, if $\varphi$ satisfies Klein-Gordon equation in the $g_{\mu\nu}$ metric then it will also satisfies it in the $q_{(A)\mu\nu}$ metric. Under the condition that the only means by which we can analyze the metric of the spacetime is through the dynamics of the scalar field, we become in all cases unable to distinguish between any of the non-equivalent metrics related by transformation (\ref{qA}).

We end our considerations with the following yet open questions. Is there a similar internal symmetry associated with the dynamics of other physical fields such as the gauge bosons or spinors? Does this spacetime ambiguity still remains in the realm of quantum physics? We will investigate these problems in a forthcoming paper.\\



\section{acknowledgements}
We would like to thank S. A. Dias for helpful comments and useful discussions. E. Goulart would like to thank FAPERJ for financial support.



\begin{thebibliography}{50}
 

\bibitem{Wig} Eugene P. Wigner, \emph{Symmetries and Reflections}, Scientific Essays, (M.I.T. Press, 1970).

\bibitem{Gross} David J. Gross, \emph{Gauge Theory-Past, Present, and Future?}, Chinese Journal of Physics	VOL. 30, NO. 7, 1992.

\bibitem{velt} M.J.G. Veltman B.Q.P.J. de Wit and G. ’t Hooft, \emph{Lie Groups in Physics} (Utrecht University, 2007).

\bibitem{Leite} J. Leite Lopes, \emph{Gauge Field Theory} (Pergamon Press, 1981).


\bibitem{wald} R. M. Wald. \emph{General Relativity}. Chicago university press, Chicago, 1984.

\bibitem{Wein} S. Weinberg, \emph{The Quantum Theory of Fields} (Cambridge University Press, 1995).
 
\bibitem{cou} R. Courant and D. Hilbert. \emph{Methods of Mathematical Physics}  (Wiley-Interscience, 1989) volume 2.

\bibitem{gordon} Gordon W. Zur Lichtfortpflanzung nach der Relativitatstheorie. Ann. Phys. Leipzig, 72 (1923), 421–456. 
 

\bibitem{nov1} Novello M; Perez Bergliaffa S E; Salim J; De Lorenci V; and Klippert R. Analog black holes in flowing dielectrics. Class. Quant. Grav., 20 (2003), 859–872. gr-qc/0201061.

\bibitem{matt} Analogue gravity, Carlos Barcelo, Stefano Liberati, Matt Visser, Living Rev.Rel.8:12,2005, gr-qc/0505065.

\bibitem{nov2} Novello M; Visser M; and Volovik G (editors). Artificial black holes. World Scientific, Singapore (2002).

\bibitem{conf} Philippe Francesco, Pierre Mathieu and David Senechal, \emph{Conformal Field Theory} (Springer, 1997)

 \bibitem{Kast} H.A. Kastrup, \emph{On the Advancements of Conformal Transformations and their Associated Symmetries in Geometry and Theoretical Physics} ....... arXiv:0808.2730v1, 2008

\bibitem{Had} Y. Choquet-Bruhat, C. de Witt-Morette, and M. Dillard-Bleick, \textit{Analysis, Manifolds and Physics} (North-Holland, New York, 1977) p. 455.

 
 

 

\end{thebibliography}
\end{document}